\documentstyle [11pt] {article}
\makeatletter
\newcommand{\doublespacing}{\let\CS=\@currsize\renewcommand{\baselinestretch}{1.4}\tiny\CS}

\newtheorem{property}{Property}

\newtheorem{definition}{Def.}

\newcommand {\tab} {\hspace*{2em}}

\date{}
\newcommand{\D}{\displaystyle}
\newcommand{\am}{a_{ij\mu}}
\newcommand{\an}{a_{ij\nu}}

\newcommand{\aml}{a_{ij\mu L}}
\newcommand{\anl}{a_{ij\nu L}}
\newcommand{\bml}{b_{ij\mu L}}
\newcommand{\bnl}{b_{ij\nu L}}
\newcommand{\amu}{a_{ij\mu U}}
\newcommand{\anu}{a_{ij\nu U}}
\newcommand{\bmu}{b_{ij\mu U}}
\newcommand{\bnu}{b_{ij\nu U}}
\newcommand{\cml}{c_{ij\mu L}}
\newcommand{\cnl}{c_{ij\nu L}}
\newcommand{\cmu}{c_{ij\mu U}}
\newcommand{\cnu}{c_{ij\nu U}}
\newcommand{\dml}{d_{ij\mu L}}
\newcommand{\dnl}{d_{ij\nu L}}
\newcommand{\dmu}{d_{ij\mu U}}
\newcommand{\dnu}{c_{ij\nu U}}
\newcommand{\x}{x_{ij}}

\begin {document}
\doublespacing

\title {\Large {\bf
Interval-Valued Intuitionistic Fuzzy Matrices}}

\author{Susanta Kumar Khan and Madhumangal Pal \thanks{e-mail: mmpalvu@gmail.com} \\
Department of Applied Mathematics with Oceanology and Computer
Programming,
\\Vidyasagar University,  Midnapore -- 721102,  \\ India.
}
\maketitle

\subsection* {\centering Abstract}
In this paper,  the interval-valued intuitionistic fuzzy matrix
(IVIFM) is introduced. The interval-valued intuitionistic fuzzy
determinant is also defined. Some fundamental operations are also
presented. The need of IVIFM is explain by an example.

\noindent {\bf Keywords:} Intuitionistic fuzzy matrix,
interval-valued intuitionistic fuzzy matrix.


\section{Introduction}
Matrices play important roles in various areas in science and
engineering. The classical matrix theory can not solve the
problems involving various types of uncertainties. That type of
problems are solved by using fuzzy matrix \cite{tho77}. Later much
works have been done by many researchers. Fuzzy matrix deals with
only membership values. These matrices can not deal non membership
values. Intuitionistic fuzzy matrices (IFMs)  introduced first
time by Khan, Shyamal and Pal \cite{pal02}. Several properties on
IFMs have been studied in \cite{kha06}. But, practically it is
difficult to measure the membership or non membership value as a
point. So, we consider the membership value as an interval and
also in the case of non membership values, it is not selected as a
point, it can be considered as an interval. Here, we introduce the
interval valued intuitionistic fuzzy matrices (IVIFMs) and
introduce some basic operators on IVIFMs. The interval-valued
intuitionistic fuzzy determinant (IVIFD) is also defined. A real
life problem on IVIFM is presented. Interpretation of some of the
operators are given with the help of this example.

\section{Definition and Preliminaries}

In this section, we first define the intuitionistic fuzzy matrix
(IFM) based on the definition of intuitionistic fuzzy sets
introduced by Atanassov \cite{ata86}. The intuitionistic fuzzy
matrices are introduced by Pal, Khan and Shyamal
\cite{pal02,kha06}.

\begin{definition}
{\bf Intuitionistic fuzzy matrix (IFM)\cite{pal02}:} An
intuitionistic fuzzy matrix (IFM) $A$ of order $m\times n$ is
defined as $A=[\x,<\am,\an>]_{m\times n}$, where $\am$ and $\an$
are called membership and non membership values of $x_{ij}$ in
$A$, which
maintaining the condition $0\le\am+\an\le 1$.\\
For simplicity, we write $A=[\x,a_{ij}]_{m\times n}$ or
simply $[a_{ij}]_{m\times n}$ where $a_{ij}=<\am,\an>$.
\end{definition}

Using the concept of intuitionistic fuzzy sets and interval
valued fuzzy sets, we define  interval-valued intuitionistic fuzzy matrices
as follows:
\begin{definition}
{\bf Interval-valued intuitionistic fuzzy matrix (IVIFM):} An interval
valued intuitionistic fuzzy matrix (IVIFM) $A$ of order $m\times n$ is
defined as $A=[\x,<\am,\an>]_{m\times n}$ where $\am$ and $\an$ are both the
subsets of $[0,1]$ which are denoted by $\am=[a_{ij\mu L},a_{ij\mu U}]$ and
$\an=[a_{ij\nu L},a_{ij\nu U}]$ which maintaining the condition
$a_{ij\mu U} + a_{ij\nu U} \le 1$  for $i=1,2,\cdots,m$ and $j=1,2,\cdots,n$.
\end{definition}

\begin{definition}
{\bf Interval-valued intuitionistic fuzzy determinant (IVIFD):} An interval
valued intuitionistic fuzzy determinant (IVIFD) function $f: M\rightarrow
F$ is a function on the set $M$ (of all $n\times n$ IVIFMs)
to the set $F$, where $F$ is the set
of elements of the form $<[a_{\mu L},a_{\mu U}],[a_{\nu L},a_{\nu U}]>$,
maintaining the condition
$0\le a_{\mu U}+a_{\nu U}\le 1$,
$0\le  a_{\mu L}\le a_{\mu U}\le 1$ and $0\le a_{\nu L}\le a_{\nu U}\le 1$
and $0\le\anl\le\anu\le 1$
such that $A\subset M$ then
$f(A)$ or $|A|$ or $det(A)$ belongs to $F$ and is given by
$$|A|= \sum_{\sigma \in S_n}\prod_{i=1}^n<[a_{i\sigma(i)\mu
L},a_{i\sigma(i)\mu U}],[ a_{i\sigma(i)\nu L},a_{i\sigma(i)\nu
U}]>$$
and $S_n$ denotes the symmetric group of all permutations of the symbols
$\{1,2,\cdots,n\}$.
\end{definition}

\begin{definition}
{\bf The adjoint IVIFM of an IVIFM:} The adjoint IVIFM of an IVIFM A of
order $n\times n$, is denoted by $adj. A$ and is defined by $adj.
A=[A_{ji}]$, where $A_{ji}$ is the determinant of the IVIFM $A$ of order
$(n-1)\times(n-1)$ formed by suppressing row $j$ and column $i$ of the
IVIFM $A$. In other words, $A_{ji}$ can be written in the form
$$\sum_{\sigma\in S_{n_{i}n_{j}}}\prod_{t\in n_{j}}<[a_{t\sigma(t)\mu
L},a_{t\sigma(t)\mu U}],[a_{t\sigma(t)\nu L},a_{t\sigma(t)\nu U}>$$ where,
$n_j=\{1,2,\ldots,n\}\backslash \{j\}$ and
$S_{n_{i}n{j}}$ is the set of all permutations of set $n_j$ over the set
$n_i$.
\end{definition}

Depending on the values of diagonal elements, the unit IVIFM are classified
into two types: $(i)\; a-unit\; IVIFM$ and $(ii)\; r-unit \;IVIFM$.

\begin{definition}
{\bf Acceptance unit IVIFM (a-unit IVIFM):} A square IVIFM is a-unit IVIFM
if all diagonal elements are $<[1,1],[0,0]>$ and all remaining elements are
$<[0,0],[1,1]>$ and it is denoted by $I_{<[0,0],[1,1]>}$.
\end{definition}

\begin{definition}
{\bf Rejection unit IVIFM (r-unit IVIFM):} A square IVIFM is a r-unit IVIFM
if all diagonal elements are $<[0,0],[1,1]>$ and all remaining elements are
$<[1,1],[0,0]>$ and it is denoted by $I_{<[1,1],[0,0]>}$.
\end{definition}

Similarly, three types of null IVIFMs are defined on its elements.
\begin{definition}
{\bf Complete null IVIFM (c-null IVIFM):} An IVIFM is a c-null IVIFM if all
the elements are $<[0,0],[0,0]>$.
\end{definition}

\begin{definition}
{\bf Acceptance null IVIFM (a-null IVIFM):} An IVIFM is a a-null IVIFM if
all the elements are $<[0,0],[1,1]>$.
\end{definition}

\begin{definition}
{\bf Rejection null IVIFM (r-null IVIFM):} An IVIFM is a r-null IVIFM if all
the elements are $<[1,1],[0,0]>$.
\end{definition}

    \subsection{Some operations on IVIFM}
Let $A=[<[\aml,\amu],[\anl,\anu]>]$ and $B=[<[\bml,\bmu],[\bnl,\bnu]>]$ be
two IVIFMs. Then,\\
(i)
$<[\aml,\amu],[\anl,\anu]>+<[\bml,\bmu],[\bnl,\bnu]>\\
\tab=<[\max(\aml,\bml),
\max(\amu,\bmu)],[\min(\anl,\bnl),\min(\anu,\bnu)]>.$\\
(ii)
$<[\aml,\amu],[\anl,\anu]>\cdot<[\bml,\bmu],[\bnl,\bnu]>\\
\tab=<[\min(\aml,\bml),
\min(\amu,\bmu)],[\max(\anl,\bnl),\max(\anu,\bnu)]>$.\\
(iii)
$A+B=[<[\max\{\aml,\bml\},
\max\{\amu,\bmu\}],[\min\{\anl,\bnl\},\min\{\anu,\bnu\}]>].$\\
(iv)
$A\cdot B=[<[\min\{\aml,\bml\},
\min\{\amu,\bmu\}],[\max\{\anl,\bnl\},\max\{\anu,\bnu\}]>].$\\
(v)
$\bar{A}=[<[\anl,\anu],[\aml,\amu]>].$ (complement of $A$)\\
(vi) $A^T=[<[a_{ji\mu L}, a_{ji\mu U}],
[a_{ji\nu L},a_{ji\nu U}]>]_{n\times m}.$ (transpose of A)\\
(vii)
$A\oplus B=[<[\aml+\bml-\aml\cdot
\bml,\amu+\bmu-\amu\cdot\bmu],\\
\tab\tab\tab\tab[\anl+\bnl-\anl\cdot\bnl,\anu+\bnu-\anu\cdot\bnu]>].$\\
(viii)
$A\odot
B=[<[\aml.\bml,\amu.\bmu],$\\
\tab\tab\tab$[\anl+\bnl-\anl.\bnl,\anu+\bnu-\anu.\bnu]>].$

\vspace{.4cm}\noindent (ix) ${\D
A@B=\Big[\Big<\Big[\frac{\aml+\bml}{2},\frac{\amu+\bmu}{2}\Big],\Big[\frac{\anl+\bnl}{2},
\frac{\anu+\bnu}{2}\Big]\Big>\Big].}$

\vspace{.4cm} \noindent (x) ${\D
A\$B=\Big[\Big<\Big[\sqrt{\aml.\bml},\sqrt{\amu.\bmu}
\Big],\Big[\sqrt{\anl.\bnl}, \sqrt{\anu.\bnu} \Big]\Big>\Big].}$

\vspace{.4cm} \noindent (xi) ${\D
A\#B=\Big[\Big<\Big[\frac{2\aml.\bml}{\aml+\bml},\frac{2\amu.\bmu}{\amu+\amu}\Big],
\Big[\frac{2\anl.\bnl}{\anl+\bnl},\frac{2\anu.\bnu}{\anu+\bnu}\Big]\Big>\Big].}$

\vspace{.4cm}\noindent (xii) ${\D A\ast
B=\Big[\Big<\Big[\frac{\aml+\bml}{2(\aml.\bml+1)},\frac{\amu+\bmu}{2(\amu.\bmu+1)}\Big],
\Big[\frac{\anl+\bnl}{2(\anl.\bnl+1)},\frac{\anu+\bnu}{2(\anu.\bnu+1)}\Big]\Big>\Big].}$

\vspace{.4cm}\noindent (xiii) $A\le B {\;\rm iff\;} \aml\le \bml,
\amu\le\bmu, \anl\ge\bnl {\;\rm and\;}
\anu\ge\bnu.$\\
(xiv)
$A=B {\;\rm iff\;} A\le B {\;\rm and\;} B\le A.$

\bigskip
In the following section, we consider a daily life problem which can be
studied using IVIFMs in better way.

\section{Need of IVIFM}
We consider a network consisting of six important cities (vertices) in a
country. They are interconnected by roads (edges). The network is shown in
Figure 1.

\bigskip
\unitlength=1.00mm
\special{em:linewidth 0.4pt}
\linethickness{0.4pt}
\begin{picture}(119.35,59.69)
\put(26.33,36.67){\circle{6.04}} \put(51.33,56.67){\circle{6.04}}
\put(86.33,36.67){\circle{6.00}} \put(66.33,6.67){\circle{6.00}}
\put(96.33,16.67){\circle{6.00}} \put(116.33,56.67){\circle{6.04}}
\put(54.33,56.67){\line(1,0){59.00}}
\put(113.00,56.34){\line(-6,-5){24.00}}
\put(51.33,53.67){\line(2,-1){32.33}}
\put(48.33,56.67){\line(-6,-5){20.67}}
\put(26.00,33.67){\line(4,-3){39.33}}
\put(51.33,53.67){\line(6,-5){42.67}}
\put(96.66,19.67){\line(3,5){20.33}}
\put(28.00,34.67){\line(4,-1){65.67}}
\put(86.33,33.67){\line(3,-5){8.67}}
\put(29.00,37.67){\line(5,1){85.00}}
\put(51.00,53.67){\line(1,-3){14.67}}
\put(29.00,36.67){\line(1,0){54.33}}
\put(26.00,36.67){\makebox(0,0)[cc]{1}}
\put(51.00,56.67){\makebox(0,0)[cc]{2}}
\put(116.33,57.00){\makebox(0,0)[cc]{3}}
\put(86.33,36.67){\makebox(0,0)[cc]{4}}
\put(96.33,17.00){\makebox(0,0)[cc]{5}}
\put(66.33,6.67){\makebox(0,0)[cc]{6}}
\put(35.00,49.34){\makebox(0,0)[cc]{10}}
\put(82.00,58.34){\makebox(0,0)[cc]{55}}
\put(108.67,34.67){\makebox(0,0)[cc]{25}}
\put(85.33,10.00){\makebox(0,0)[cc]{30}}
\put(97.00,34.67){\makebox(0,0)[cc]{10}}
\put(46.00,16.00){\makebox(0,0)[cc]{10}}
\put(51.33,27.34){\makebox(0,0)[cc]{20}}
\put(59.67,18.34){\makebox(0,0)[cc]{31}}
\put(64.67,34.67){\makebox(0,0)[cc]{30}}
\put(78.67,50.00){\makebox(0,0)[cc]{15}}
\put(94.67,44.34){\makebox(0,0)[cc]{70}}
\put(78.33,25.67){\makebox(0,0)[cc]{10}}
\put(78.33,42.67){\makebox(0,0)[cc]{40}}
\put(73.67,32.34){\makebox(0,0)[cc]{18}}
\put(93.33,25.67){\makebox(0,0)[cc]{5}}
\put(93.67,15.00){\line(-3,-1){25.00}}
\put(86.33,33.67){\line(-3,-4){18.67}}
\put(115.00,54.34){\line(-1,-1){46.33}}
\put(71.67,1.33){\makebox(0,0)[cc]{{\bf Figure 1:} A network.}}
\end{picture}

The number adjacent to an edge represents the distance between the cities
(vertices). The above network can be represented with the help of a
classical matrix $A=[a_{ij}]$, $i,j=1,2,\ldots,n$, where, $n$ is the total
number of nodes. The $ij$th element $a_{ij}$ of $A$ is defined as

$$a_{ij}=\left\{\begin{array}{cl}
0, &{\;\rm if\;}i=j\\
\infty, &{\;\rm the\; vertices \;}i{\;\rm and\;}j{\;\rm are\; not
\;directly\; connected\; by\; an\; edge}\\
w_{ij}, &w_{ij}{\;\rm is\; the\; distance\; of\; the\; road\; connecting\;
}i{\; \rm and\;} j.
\end{array}\right.$$

Thus the adjacent matrix of the network of Figure 1 is
\newpage
$$\begin{array}{c c c c c c }
\tab 1\;\;&2\;\;&3\;&4\;&5\;\;&6
\end{array}$$
$$\begin{array}{c}
1\\2\\3\\4\\5\\6
\end{array}
\left[\begin{array}{c c c c c c c}
0&10&15&30&20&10\\
10&0&55&40&18&30\\
15&55&0&70&25&10\\
30&40&70&0&5&10\\
20&18&25&5&0&30\\
10&31&10&10&30&0
\end{array}\right]$$

Since the distance between two vertices are known, precisely, so the above
matrix is obviously a classical matrix.
Generally, the distance between two cities are crisp value, so the
corresponding matrix is crisp matrix.

Now, we consider the crowdness of the roads connecting cities. It is clear
that the crowdness of a road obviously, is a fuzzy quantity. The amount of
crowdness depends on the decision makers mentality, habits, natures, etc.
i.e., completely depends on the decision maker. The measurement of crowdness
as a point is a difficult task for the decision maker. So, here we consider
the amount of crowdness as an interval instead of a point. Similarly, the
loneliness is also considered as an interval. The crowdness and loneliness
of a network can not be represented as a crisp matrix, it can be
represented appropriately by a matrix which we designate by
interval-valued intuitionistic fuzzy matrices (IVIFMs).

For illustration, we
consider the crowdness and loneliness of the road $(i,j)$ connecting the
places $i$ and $j$ as follows:

\begin{center}
\begin{tabular}{|l|cccccccc|}
\hline Roads & (1,2) & (1,3) & (1,4) & (1,5) &  (1,6) &  (2,3) &
(2,4) & (2,5)\\
Crowdness & [.1,.3] & [.2,.4] & [.3,.4] & [.2,.4]& [.3,.6]
&[.7,.8]&[.3,.5]&[.3,.4]\\
Loneliness
&[.2,.5]&[.1,.5]&[.5,.6]&[.4,.5]&[.2,.3]&[0,.1]&[.4,.5]&[.4,.6]\\
\hline
Roads & (2,6) & (3,4) & (3,5) &  (3,6) &  (4,5) &  (4,6)&(5,6)& \\
Crowdness &[.2,.3]&[.5,.6]&[.3,.5]&[.3,.6]&[.4,.6]&[.2,.4]&[.3,.5]&\\
Loneliness&[.4,.5]&[.2,.3]&[.2,.3]&[.2,.3]& [.3,.4]&[.3,.5]
&[.2,.4]&\\
\hline
\end{tabular}

\vspace{.5 cm} {\bf Table 1:} The crowdness and loneliness of the
network of Figure 1.
\end{center}

\bigskip
The matrix representation of the traffic crowdness and loneliness of
the network of Figure 1 is shown in the following IVIFM.

\newpage
\setlength{\arraycolsep}{.5mm}

$\begin{array}{c c c c c c } \tab\tab\;\;
1\tab&\tab\tab2\tab&\tab\;\;\;\;3\tab&\tab\;\;\;\;4\tab&\tab\;\;\;\;\;\;5\tab&\tab\;\;\;\;\;6\tab
\end{array}$
{\scriptsize
$$\begin{array}{c}
1\\2\\3\\4\\5\\6
\end{array}
\left[\begin{array}{c c c c c c}
<[0,0],[1,1]>&<[.1,.3],[.2,.5]>&<[.2,.4],[.1,.5]>&<[.3,.4],[.5,.6]>&
<[.2,.4],[.4,.5]>&<[.3,.6],[.2,.3]>\\
<[.1,.3],[.2,.5]>&<[0,0],[1,1]>&<[.7,.8],[0,.1]>&<[.3,.5],[.4,.5]>&
<[.3,.4],[.4,.6]>&<[.2,.3],[.4,.5]>\\
<[.2,.4],[.1,.5]>&<[.7,.8],[0,.1]>&<[0,0],[1,1]>&<[.5,.6],[.2,.3]>&
<[.3,.5],[.2,.3]>&<[.3,.6],[.2,.3]>\\
<[.3,.4],[.5,.6]>&<[.3,.5],[.4,.5]>&<[.5,.6],[.2,.3]>&<[0,0],[1,1]>&
<[.4,.6],[.3,.4]>&<[.2,.4],[.3,.5]>\\
<[.2,.4],[.4,.5]>&<[.3,.4],[.4,.6]>&<[.3,.5],[.2,.3]>&<[.4,.6],[.3,.4]>&
<[0,0],[1,1]>&<[.3,.5],[.2,.4]>\\
<[.3,.6],[.2,.3]>&<[.2,.3],[.4,.5]>&<[.3,.6],[.2,.3]>&<[.2,.4],[.3,.5]>&
<[.3,.5],[.2,.4]>&<[0,0],[1,1]>\\
\end{array}
\right]$$}

To explain the meaning of the operators defined earlier we consider  two
IVIFMs $A$ and $B$. Let $A$ and $B$ represent respectively the crowdness
and the loneliness of the network at two time instances $t$ and $t'$. Now,
the IVIFM $A+B$ represents the maximum amount of traffic crowdness
and minimum amount of loneliness of the network between the
time instances $t$ and $t'$. $A.B$ represents the
minimum amount of traffic crowdness and maximum amount of loneliness of the
network. $\bar{A}$ matrix represents the loneliness  and
crowdness of the network. $A@B, A\$B$ and $A\#B$ reveals the arithmetic
mean, geometric mean and harmonic mean of the crowdness and loneliness in
between the two time instances $t$ and $t'$ of the network .

\begin{center}
\unitlength=1.00mm
\special{em:linewidth 0.4pt}
\linethickness{0.4pt}
\begin{picture}(67.00,25.67)
\put(8.33,4.34){\line(5,3){30.67}}
\put(39.00,22.67){\line(4,-3){24.33}}
\put(63.33,4.34){\line(-1,0){55.00}}
\put(5.00,3.67){\makebox(0,0)[cc]{1}}
\put(38.66,25.67){\makebox(0,0)[cc]{2}}
\put(67.00,4.34){\makebox(0,0)[cc]{3}}
\put(22.00,16.67){\makebox(0,0)[rb]{$\langle[.1,.3],[.2,.5]\rangle$}}
\put(54.00,16.67){\makebox(0,0)[lc]{$\langle[.7,.8],[0,.1]\rangle$}}
\put(35.66,0.34){\makebox(0,0)[cc]{$\langle[.2,.4],[.1,.5]\rangle$}}
\end{picture}

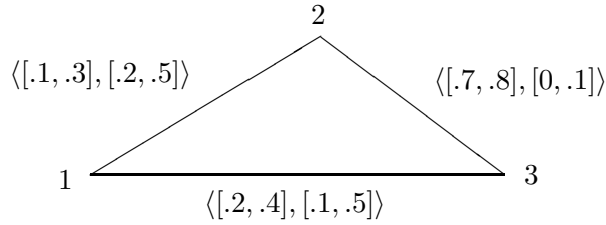
{\bf Figure 2:}
\end{center}

\bigskip

\begin{center}
\unitlength=1.00mm
\special{em:linewidth 0.4pt}
\linethickness{0.4pt}
\begin{picture}(76.67,29.00)
\put(18.00,7.67){\line(5,3){30.67}}
\put(48.67,26.00){\line(4,-3){24.33}}
\put(73.00,7.67){\line(-1,0){55.00}}
\put(14.67,7.00){\makebox(0,0)[cc]{1}}
\put(48.33,29.00){\makebox(0,0)[cc]{2}}
\put(76.67,7.67){\makebox(0,0)[cc]{3}}
\put(31.67,20.00){\makebox(0,0)[rb]{$\langle[.2,.4],[..4.5]\rangle$}}
\put(63.67,20.00){\makebox(0,0)[lc]{$\langle[.2,.4],[.3,.5]\rangle$}}
\put(45.33,3.67){\makebox(0,0)[cc]{$\langle[.3,.6],[.2,.3]\rangle$}}
\end{picture}

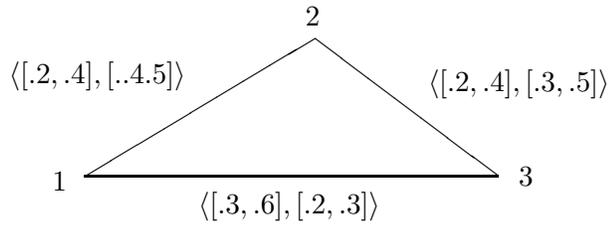
{\bf Figure 3:}
\end{center}

\medskip
To illustrate the operators $A.B$, $A+B$ and $|A|$, we consider a network
consisting three vertices and three edges. The crowdness and loneliness of
the network are observed at two different time instances $t$ and $t'$. The
matrices $A_t$ and $A_{t'}$ represent the status of the network at $t$
(Figure 2) and at $t'$ (Figure 3). The number adjacent to the sides
represents the crowdness and loneliness of the roads at two different
instances of the same network. $A_t$ and $A_{t'}$ be the matrix
representation of crowdness and loneliness at time $t$ and $t'$ respectively,

$${\rm Let \;\;}A_t=\left[\begin{array}{ccc}
<[0,0],[1,1]>&<[.1,.3],[.2,.5]>&<[.2,.4],[.1,.5]>\\
<[.1,.3],[.2,.5]>&<[0,0],[1,1]>&<[.7,.8],[0,.1]>\\
<[.2,.4],[.1,.5]>&<[.7,.8],[0,.1]>&<[0,0],[1,1]>
\end{array}
\right]$$

$${\rm and \;\;}A_{t'}=\left[\begin{array}{ccc}
<[0,0],[1,1]>&<[.2,.4],[.4,.5]>&<[.3,.6],[.2,.3]>\\
<[.2,.4],[.1,.5]>&<[0,0],[1,1]>&<[.2,.4],[.3,.5]>\\
<[.3,.6],[.2,.3]>&<[.2,.4],[.3,.5]>&<[0,0],[1,1]>
\end{array}
\right].$$

$${\rm So,\;\;} A_t.A_{t'}=\left[\begin{array}{ccc}
<[0,0],[1,1]>&<[.1,.3],[.4,.5]>&<[.2,.4],[.2,.5]>\\
<[.1,.3],[.4,.5]>&<[0,0],[1,1]>&<[.2,.4],[.3,.5]>\\
<[.2,.4],[.2,.5]>&<[.2,.4],[.3,.5]>&<[0,0],[1,1]>
\end{array}
\right]$$

$${\rm and,\;\;} A_t+A_{t'}=\left[\begin{array}{ccc}
<[0,0],[1,1]>&<[.2,.4],[.2,.5]>&<[.3,.6],[.1,.3]>\\
<[.2,.4],[.2,.5]>&<[0,0],[1,1]>&<[.7,.8],[ 0,.1]>\\
<[.3,.6],[.1,.3]>&<[.7,.8],[ 0,.1]>&<[0,0],[1,1]>
\end{array}
\right].$$

$$\left.\begin{array}{rl}
|A_t|=&<[0,0],[1,1]>\{<[0,0],[1,1]><[0,0],[1,1]>+<[.7,.8],[.0,.1]><[.7,.8],[0,1]>\}\\
&+<[.1,.3],[.2,.5]>\{<[.7,.8],[0,.1]><[.2,.4],[.1,.5]>+<.1,.3],[.2,.5]><[0,0],[1,1]>\}\\
&+<[.2,.4],[.1,.5]>\{<[.1,.3],[.2,.5]><.7,.8],[0,.1]>+<[0,0],[1,1]><[.2,.4],[.1,.5]>\}\\
=&<[0,0],[1,1]>\{<[0,0],[1,1]>+<[.7,.8],[0,.1]>\}\\
&+<[.1,.3],[.2,.5]>\{<[.2,.4],[.1,.5]>+<[0,0],[1,1]>\}\\
&+<[.2,.4],[.1,.5]>\{<[.1,.3],[.2,.5]>+<[0,0],[1,1]>\}\\
=&<[0,0],[1,1]><[.7,.8],[0,.1]>\\
&+<[.1,.3],[.2,.5]><[.2,.4],[.1,.5]>\\
&+<[.2,.4],[.1,.5]><[.1,.3],[.2,.5]>\\
=&<[0,0],[1,1]>+<[.1,.3],[.2,.5]>+<[.1,.3],[.2,.5]>\\
=&<[.1,.3],[.2,.5]>
\end{array}
\right.$$

It may be noted that if the $ij$-th element of the IVIFM
$A_t$ is $<[0,0],[1,1]>$ then it indicates that the road $(i,j)$ is fully
lonely (not crowd), but, if it is $<[1,1],[0,0]>$ then the road $(i,j)$ is
fully crowd or blocked.

\section{Properties of IVIFMs}
In this section some properties of IVIFMs are presented.\\
IVIFMs satisfy the commutative and associative properties over the operators
$+,.,\oplus$, and $\odot$. The operator `.' is distributed over `$+$' in
left and right but the left and right distribution laws  do not hold for the
operators $\oplus$ and $\odot$.\\
(1) $A+B=B+A$\\
(2) $A+(B+C)=(A+B)+C$\\
(3) $A.B=B.A$\\
(4) $A.(B.C)=(A.B).C$\\
(5) (i)$A.(B+C)=A.B+A.C$\\
\tab(ii) $(B+C).A=B.A+C.A$\\
(6) $A\oplus B=B\oplus A$\\
(7) $A\oplus (B\oplus C)=(A\oplus B)\oplus C$\\
(8) $A\odot B=B\odot A$\\
(9) $A\odot (B\odot C)=(A\odot B)\odot C$\\
(10)(i) $A\odot(B\oplus C)\neq (A\odot B)\oplus (A\odot C)$\\
\tab(ii) $(B\oplus C)\odot A\neq (B\odot A)\oplus (C\odot A)$

\noindent {\bf Proof of (i):}
Let $A=[<[\aml,\amu],[\anl,\anu]>]$,\\
$B=[<[\bml,\bmu],[\bnl,\bnu]>]$\\
and $C=[<[\cml,\cmu],[\cnl,\cnu]>]$.\\
So,
$B\oplus
C=[<[\bml+\cml-\bml.\cml,\bmu+\cmu-\bmu.\cmu],[\bnl.\cnl,\bnu.\cnu]>]$\\
and $A\odot(B\oplus C)=[<[\aml.(\bml+\cml-\bml.\cml),
\amu(\bmu+\cmu-\bmu.\cmu],[\anl+\bnl.\cnl-\anl.\bnl.\cnl,\amu+\bmu.\cmu-\amu.\bmu.\cmu]>]$.\\
$A\odot
B=[<[\aml.\bml,\amu.\bmu],[\anl+\bnl-\anl.\bnl,\anu+\bnu-\anu.\bnu]>],$\\
$A\odot
C=[<[\aml.\cml,\amu.\cmu],[\anl+\cnl-\anl.\cnl,\anu+\cnu-\anu.\cnu]>].$\\
Now, $(A\odot B)\oplus(A\odot
C)=[<[\aml(\bml+\cml)-\aml^2.\bml.\cml,\amu(\bmu+\cmu)-\amu^2.\bmu.\cmu],
[(\anl+\bnl-\anl.\bnl).(\anl+\cnl-\anl.\cnl),(\anu+\bnu-\anu.\bnu).\\(\anu+\cnu-\anu.\cnu)]>].$\\
So, $A\odot(B\oplus C)\neq(A\odot B)\oplus(A\odot C).$

\begin{property}
Let $A$ be an IVIFM of any order then, $A+A=A.$
\end{property}
{\bf Proof:} Let $A=[<[\aml,\amu],[\anl,\anu]>]$\\
Then $A+A=[<[{\rm max}(\aml,\aml),{\rm max}(\amu,\amu)],[{\rm
min}(\anl,\anl),
{\min}(\anu,\anu)]>]\\
\tab\tab=[<[\aml,\amu],[\anl,\anu]>]$\\
\tab\tab=$A$.

\begin{property}
 If $A$ be an IVIFM of any order then, $A+I_{<[0,0],[0,0]>}\ge A$
where, $I_{<[0,0],[0,0]>}$ is the null IVIFM of same order.
\end{property}
{\bf Proof:} Let $A=[<[\aml,\amu],[\anl,\anu]>]$\\
and $I_{<[0,0],[0,0]>}=<[0,0],[0,0]>.$\\
Then, $A+I_{<[0,0],[0,0]>}=[<[{\rm max}(\aml,0),{\rm max}(\amu,0)],
[{\rm min}(\anl,0),{\rm min}(\anu,0)]>]\\
\tab\tab\tab\tab\tab=[<[\aml,\amu],[0,0]>]$\\
Therefore, $A+I_{<[0,0],[0,0]>}\ge A.$

Some more properties on determinant and adjoint of IVIFM are presented below.

\begin{property} Like classical matrices the determinant value of an IVIFM and
its transpose are equal. If $A$ be a square IVIFM then
$|A|=|A^T|$.
\end{property}
{\bf Proof:} Let $A=[<[\aml,\amu],[\anl,\anu]>].$\\
Then $A^T=B=[<[\bml,\bmu],[\bnl,\bnu]>]$\\
$\tab\tab=[<[a_{ji\mu L},a_{ji\mu U}],[a_{ji\nu L},a_{ji\nu U}]>].$\\
Now, $$|B|=\sum_{\sigma\in S_n}<[b_{1\sigma(1)\mu L},b_{1\sigma(1)\mu U}],
[b_{1\sigma(1)\nu L},b_{1\sigma(1)\nu U}]><[b_{2\sigma(2)\mu L},
b_{2\sigma(2)\mu U}],[b_{2\sigma(2)\nu L},b_{2\sigma(2)\nu U}]>\ldots$$
$\tab\tab\,<[b_{n\sigma(n)\mu L},b_{n\sigma(n)\mu U}],[b_{n\sigma(n)\nu
L},b_{n\sigma(n)\nu U}]>$
$$\;=\sum_{\sigma\in S_n}<[a_{\sigma(1)1\mu L},a_{\sigma(1)1\mu U}],
[a_{\sigma(1)1\nu L},a_{\sigma(1)1\nu U}]><[a_{\sigma(2)2\mu L},
a_{\sigma(2)2\mu U}],[a_{\sigma(2)2\nu L},a_{\sigma(2)2\nu
U}]>\ldots$$ $\tab\tab\,<[a_{\sigma(n)n\mu L},a_{\sigma(n)n\mu
U}],[a_{\sigma(n)n\nu L},a_{\sigma(n)n\nu U}]>.$

 Let $\psi$ be the
permutation of $\{1,2\ldots n\}$ such that $\psi\sigma=I$, the
identity permutation. Then $\psi={\sigma}^{-1}$. As
$\sigma$ runs over the whole set of permutations, so does $\psi$.\\
Let $\sigma(i)=j,\;i={\sigma}^{-1}(j)=\psi(j).$

 Therefore,
$a_{\sigma(i)i\mu L}=a_{j\psi(j)\mu L},\;a_{\sigma(i)i\mu U}=
a_{j\psi(j)\mu U},\;a_{\sigma(i)i\nu L}=a_{j\psi(j)\nu
L},\;a_{\sigma(i)i\nu U}=a_{j\psi(j)\nu U}$ for all $i,j$.\\
As $i$ runs over the set $\{1,2,\ldots,n\}$, $j$ does so.\\
Now, $<[a_{\sigma(1)1\mu L},a_{\sigma(1)1\mu U}],
[a_{\sigma(1)1\nu L},a_{\sigma(1)1\nu U}]><[a_{\sigma(2)2\mu L},
a_{\sigma(2)2\mu U}],[a_{\sigma(2)2\nu L},a_{\sigma(2)2\nu U}]>\ldots$\\
$\tab\tab\,<[a_{\sigma(n)n\mu L},a_{\sigma(n)n\mu
U}],[a_{\sigma(n)n\nu L},a_{\sigma(n)n\nu U}]>$\\
=$<[a_{1\psi(1)\mu L},a_{1\psi(1)\mu U}],
[a_{1\psi(1)\nu L},a_{1\psi(1)\nu U}]><[a_{2\psi(2)\mu L},
a_{2\psi(2)\mu U}],[a_{2\psi(2)\nu L},a_{2\psi(2)\nu U}]>\ldots$\\
$\tab\tab\,<[a_{n\psi(n)\mu L},a_{n\psi(n)\mu U}],[a_{n\psi(n)\nu
L},a_{n\psi(n)\nu U}]>.$ \\
Therefore,$$|B|=\sum_{\sigma\in S_n}<[a_{\sigma(1)1\mu L},a_{\sigma(1)1\mu
U}],
[a_{\sigma(1)1\nu L},a_{\sigma(1)1\nu U}]><[a_{\sigma(2)2\mu L},
a_{\sigma(2)2\mu U}],[a_{\sigma(2)2\nu L},a_{\sigma(2)2\nu U}]>\ldots$$
$\tab\tab\,<[a_{\sigma(n)n\mu L},a_{\sigma(n)n\mu U}],[a_{\sigma(n)n\nu
L},a_{\sigma(n)n\nu U}]>$\\
$$=\sum_{\psi\in S_n}<[a_{1\psi(1)\mu L},a_{1\psi(1)\mu U}],
[a_{1\psi(1)\nu L},a_{1\psi(1)\nu U}]><[a_{2\psi(2)\mu L},
a_{2\psi(2)\mu U}],[a_{2\psi(2)\nu L},a_{2\psi(2)\nu U}]>\ldots$$
$\tab\tab\,<[a_{n\psi(n)\mu L},a_{n\psi(n)\mu U}],[a_{n\psi(n)\nu
L},a_{n\psi(n)\nu U}]>$ \\
$=|A|.$

\begin{property}
 If $A$ and $B$ be two square IVIFMs and $A\le B$, then,
$adj.\,A\le adj.\,B$.
\end{property}
{\bf Proof:}
Let, $C=[<[\cml,\cmu],[\cnl,\cnu]>]=adj.\,A,$\\
$D=[<[\dml,\dmu],[\dnl,\dnu]>]=adj.\,B$\\
${\rm where,\;}<[\cml,\cmu],[\cnl,\cnu]>\; =\sum_{\sigma\in
S_{n_{i}n_{j}}}\prod_{t\in n_{j}}<[a_{t\sigma(t)\mu
L},a_{t\sigma(t)\mu U}],[a_{t\sigma(t)\nu L},a_{t\sigma(t)\nu U}>$
${\rm and\;}<[\dml,\dmu],[\dnl,\dnu]>\; =\sum_{\sigma\in
S_{n_{i}n_{j}}}\prod_{t\in n_{j}}<[b_{t\sigma(t)\mu
L},b_{t\sigma(t)\mu U}],[b_{t\sigma(t)\nu L},b_{t\sigma(t)\nu
U}>.$ It is clear that
$<[\cml,\cmu],[\cnl,\cnu]>\;\le\;<[\dml,\dmu],[\dnl,\dnu]>.$\\
Since, $a_{t\sigma(t)\mu L}\;\le\;b_{t\sigma(t)\mu L},\;a_{t\sigma(t)\mu
U}\;\le\;b_{t\sigma(t)\mu U},\;a_{t\sigma(t)\nu L}\;\ge\;b_{t\sigma(t)\nu
L},\;{\rm and\;}a_{t\sigma(t)\nu U}\;\ge\;b_{t\sigma(t)\nu U}$\\
for all $t\neq j,\;\sigma(t)\neq\sigma(j).$\\
Therefore $C\le D$, i.e., $adj.\,A\le adj.\,B$.

\begin{property}
For a square IVIFM A, $adj.\,(A^T)=(adj.\,A)^T.$
\end{property}
{\bf Proof:} Let $B=adj.\,A,\;C=adj.\,A^T$.\\
Therefore, $<[\bml,\bmu],[\bnl,\bnu]>\;=\sum_{\sigma\in
S_{n_{j}n_{i}}}\prod_{t\in n_{i}}<[a_{t\sigma(t)\mu
L},a_{t\sigma(t)\mu U}],[a_{t\sigma(t)\nu L},a_{t\sigma(t)\nu U}>$
${\rm and\;}<[\cml,\cmu],[\cnl,\cnu]>\;=\sum_{\sigma\in
S_{n_{i}n_{j}}}\prod_{t\in n_{j}}<[a_{t\sigma(t)\mu
L},a_{t\sigma(t)\mu U}],[a_{t\sigma(t)\nu L},a_{t\sigma(t)\nu U}>$
$\hspace{6.5cm}=<[\bml,\bmu],[\bnl,\bnu]>.$\\
Therefore, $adj.\,(A^T)=(adj.\,A)^T.$

The following result is not valid for classical matrices, though it is true
for IVIFM.

\begin{property}
For an IVIFM $A$, $|A|=|adj.\,A|.$
\end{property}
{\bf Proof:}
$adj.\,A=[<[A_{ij\mu L},A_{ij\mu U}],[A_{ij\nu L},A_{ij\nu U}]>].$\\
where, $<[A_{ij\mu L},A_{ij\mu U}],[A_{ij\nu L},A_{ij\nu U}]>$ is
the cofactor of the element $<[a_{ij\mu L},a_{ij\mu U}],[a_{ij\nu
L},a_{ij\nu U}]>$ in the IVIFM $A$.

 Therefore,
$|adj.\,A|=\sum_{\sigma\in S_n}<[A_{1\sigma(1)\mu
L},A_{1\sigma(1)\mu U}], [A_{1\sigma(1)\nu L},A_{1\sigma(1)\nu
U}]>\\\tab\tab\tab\;\tab\tab<[A_{2\sigma(2)\mu L},
A_{2\sigma(2)\mu U}],[A_{2\sigma(2)\nu
L},A_{2\sigma(2)\nu U}]>$\\
$\tab\tab\tab\;\tab\tab\ldots<[A_{n\sigma(n)\mu
L},A_{n\sigma(n)\mu
U}],[A_{n\sigma(n)\nu L},A_{n\sigma(n)\nu U}]>$\\
$=\sum_{\sigma\in S_n}\prod_{i=1}^n<[A_{i\sigma(i)\mu
L},A_{i\sigma(i)\mu U}],[A_{i\sigma(i)\nu L},A_{i\sigma(i)\nu
U}]>$\\
$=\sum_{\sigma\in S_n}\left[\prod_{i=1}^n\left(\sum_{\theta\in
S_{n_in_{\sigma(i)}}}\prod_{t\in n_i}<[a_{t\theta(t)\mu
L},a_{t\theta(t)\mu U}],[a_{t\theta(t)\nu L},a_{t\theta(t)\nu
U}]>\right)\right]$\\
$=\sum_{\sigma\in S_n}\left[\left(\prod_{t\in
n_1}<[a_{t{\theta}_1(t)\mu L},a_{t{\theta}_1(t)\mu
U}],[a_{t{\theta}_1(t)\nu L},a_{t{\theta}_1(t)\nu
U}]>\right)\right.\left(\prod_{t\in n_2}<[a_{t{\theta}_2(t)\mu
L},a_{t{\theta}_2(t)\mu U}],\right.$\\
$\tab\tab\tab\;\;[a_{t{\theta}_2(t)\nu L},a_{t{\theta}_2(t)\nu
U}]>\Big)\ldots\left. \left(\prod_{t\in n_n}<[a_{t{\theta}_n(t)\mu
L},a_{t{\theta}_n(t)\mu U}],[a_{t{\theta}_n(t)\nu
L},a_{t{\theta}_n(t)\nu
U}]>\right)\right]\hspace{1.4cm}$\\
\hspace{5cm}$\Big($For some $\theta_1\in S_{n_1n_{\sigma(1)}},
\theta_2\in S_{n_2n_{\sigma(2)}},\ldots, \theta_n\in
S_{n_1n_{\sigma(n)}}\Big)$\\
$=\sum_{\sigma\in S_n}[(<[a_{2{\theta}_1(2)\mu
L},a_{2{\theta}_1(2)\mu U}],[a_{2{\theta}_1(2)\nu
L},a_{2{\theta}_1(2)\nu U}]> <[a_{3{\theta}_1(3)\mu
L},a_{3{\theta}_1(3)\mu U}],[a_{3{\theta}_1(3)\nu
L},a_{3{\theta}_1(3)\nu U}]>$\\
$\ldots<[a_{n{\theta}_1(n)\mu L},a_{n{\theta}_1(n)\mu
U}],[a_{n{\theta}_1(n)\nu L},a_{n{\theta}_1(n)\nu
U}]>)(<[a_{1{\theta}_2(1)\mu L},a_{1{\theta}_2(1)\mu
U}],[a_{1{\theta}_2(1)\nu L},a_{1{\theta}_2(1)\nu U}]>$\\
$ <[a_{3{\theta}_2(3)\mu L},a_{3{\theta}_2(3)\mu
U}],[a_{3{\theta}_2(3)\nu L},a_{3{\theta}_2(3)\nu U}]>\ldots
<[a_{n{\theta}_2(n)\mu L},a_{n{\theta}_2(n)\mu
U}],[a_{n{\theta}_2(n)\nu L},a_{n{\theta}_2(n)\nu U}]>)$\\
$\ldots \tab\ldots\tab\ldots\tab\ldots\tab\ldots\tab \ldots \tab
\ldots\tab\ldots\tab\ldots\tab\ldots\tab\ldots
\tab\ldots\tab\ldots$\\
$(<[a_{1{\theta}_n(1)\mu L},a_{1{\theta}_n(1)\mu
U}],[a_{1{\theta}_n(1)\nu L},a_{1{\theta}_n(1)\nu U}]>
<[a_{2{\theta}_n(2)\mu L},a_{2{\theta}_n(2)\mu
U}],[a_{2{\theta}_n(2)\nu L},a_{2{\theta}_n(2)\nu U}]>\ldots$\\
$<[a_{(n-1){\theta}_n(n-1)\mu L},a_{(n-1){\theta}_n(n-1)\mu
U}],[a_{(n-1){\theta}_n(n-1)\nu L},a_{(n-1){\theta}_n(n-1)\nu
U}]>)]\hspace{5cm}$\\
$=\sum_{\sigma\in S_n}[(<[a_{1{\theta}_2(1)\mu
L},a_{1{\theta}_2(1)\mu U}],[a_{1{\theta}_2(1)\nu
L},a_{1{\theta}_2(1)\nu U}]> <[a_{1{\theta}_3(1)\mu
L},a_{1{\theta}_3(1)\mu U}],[a_{1{\theta}_3(1)\nu
L},a_{1{\theta}_3(1)\nu U}]>$\\
$\ldots<[a_{1{\theta}_n(1)\mu L},a_{1{\theta}_n(1)\mu
U}],[a_{1{\theta}_n(1)\nu L},a_{1{\theta}_n(1)\nu
U}]>)(<[a_{2{\theta}_1(2)\mu L},a_{2{\theta}_1(2)\mu
U}],[a_{2{\theta}_1(2)\nu L},a_{2{\theta}_1(2)\nu U}]>$\\
$<[a_{2{\theta}_3(2)\mu L},a_{2{\theta}_3(2)\mu
U}],[a_{2{\theta}_3(2)\nu L},a_{2{\theta}_3(2)\nu
U}]>\ldots<[a_{2{\theta}_n(2)\mu L},a_{2{\theta}_n(2)\mu
U}],[a_{2{\theta}_n(2)\nu L},a_{2{\theta}_n(2)\nu U}]>)$\\
$\ldots \tab\ldots\tab\ldots\tab\ldots\tab\ldots\tab \ldots \tab
\ldots\tab\ldots\tab\ldots\tab\ldots\tab\ldots
\tab\ldots\tab\ldots$\\
$(<[a_{n{\theta}_1(n)\mu L},a_{n{\theta}_1(n)\mu
U}],[a_{n{\theta}_1(n)\nu L},a_{n{\theta}_1(n)\nu
U}]><[a_{n{\theta}_2(n)\mu L},a_{n{\theta}_2(n)\mu
U}],[a_{n{\theta}_2(n)\nu L},a_{n{\theta}_2(n)\nu U}]>$\\
$\ldots <[a_{n{\theta}_{(n-1)}(n)\mu L},a_{n{\theta}_{(n-1)}(n)\mu
U}],[a_{n{\theta}_{(n-1)}(n)\nu L},a_{n{\theta}_{(n-1)}(n)\nu
U}]>)]\hspace{5cm}$\\
$=\sum_{\sigma\in S_n}[<[a_{1{\theta}_{f_1}(1)\mu
L},a_{1{\theta}_{f_1}(1)\mu U}],[a_{1{\theta}_{f_1}(1)\nu
L},a_{1{\theta}_{f_1}(1)\nu U}]>\\
\tab\tab<[a_{2{\theta}_{f_2}(2)\mu L},a_{2{\theta}_{f_2}(2)\mu
U}],[a_{2{\theta}_{f_2}(2)\nu
L},a_{2{\theta}_{f_2}(2)\nu U}]>$\\
$\tab\tab\ldots<[a_{n{\theta}_{f_n}(n)\mu
L},a_{n{\theta}_{f_n}(n)\mu U}],[a_{n{\theta}_{f_n}(n)\nu
L},a_{n{\theta}_{f_n}(n)\nu U}]>]\hspace{6cm}$\\ where,
$f_{\hat{\theta}}\in\{1,2,\ldots,n\}\backslash \{\hat{\theta}\},
\hat{\theta}=1,2,\ldots,n$.

But since,
$<[a_{{\hat{\theta}}{\theta}_{f_{\hat{\theta}}}({\hat{\theta}})\mu
L},a_{{\hat{\theta}}{\theta}_{f_{\hat{\theta}}}({\hat{\theta}})\mu
U}],[a_{{\hat{\theta}}{\theta}_{f_{\hat{\theta}}}({\hat{\theta}})\nu
L},a_{{\hat{\theta}}{\theta}_{f_{\hat{\theta}}}({\hat{\theta}})\nu
U}]>\\ \tab\tab\tab\tab =<[a_{n\sigma(n)\mu L},a_{n\sigma(n)\mu
U}],[a_{n\sigma(n)\nu L},a_{n\sigma(n)\nu U}]>$.

Therefore, $|adj.\,A|=\sum_{\sigma\in S_n}<[a_{1\sigma(1)\mu
L},a_{1\sigma(1)\mu
U}],[a_{1\sigma(1)\nu L},a_{1\sigma(1)\nu U}]>$\\
\tab\tab\tab\tab\tab $<[a_{2\sigma(2)\mu L}, a_{2\sigma(2)\mu
U}],[a_{2\sigma(2)\nu
L},a_{2\sigma(2)\nu U}]>\ldots$\\
\tab\tab\tab\tab\tab $<[ a_{n\sigma(n)\mu L},a_{n\sigma(n)\mu
U}],[a_{n\sigma(n)\nu
L},a_{n\sigma(n)\nu U}]>$\\
\tab\tab\tab\tab\tab$=|A|.$

\end{document}